\documentclass[aps,pra,preprint]{revtex4}
\usepackage{graphicx}
\usepackage{latexsym}
\usepackage{amssymb}
\begin{document}
\baselineskip=24pt
\title{
Density-to-potential map in time-independent excited-state 
density-functional theory
}
\author{Prasanjit Samal}
\email{Fax:+91-512-259 0914 ; E-mail: dilu@iitk.ac.in}
\author{Manoj K. Harbola}
\email{Fax:+91-512-259 0914 ; E-mail: mkh@iitk.ac.in}
\affiliation{Department of Physics, Indian Institute of Technology,
 Kanpur U.P. 208016, India}
\author{ A. Holas}
\affiliation{Institute of Physical Chemistry of the Polish Academy of
Sciences, 44/52 Kasprzaka, 01-224 Warsaw, Poland }
\begin{abstract}
In light of the recent work by Sahni {\em et al.}, Harbola, and
Gaudoin and Burke, the question of mapping from an excited-state
density of a many-electron interacting system to the potential of
the related non-interacting system is analyzed. To do so, we investigate
the Levy-Nagy criterion quantitatively for several excited-states. Our work
indicates that Levy-Nagy criterion may fix the density to potential map
uniquely. 
\end{abstract}
\maketitle
\newpage
The question of whether there exists a mapping from an excited-state
density $\rho(\bf r)$ to a potential $v(\bf r)$ is central to
performing density-functional calculations for excited states. The issue
has been addressed recently in a series of papers by Sahni et al.
\cite{smss}, Harbola \cite{harb}, and Gaudoin and Burke
\cite{gb}. In the work of \cite{smss} and \cite{harb}, it was shown that a
given ground- or excited-state density can be generated as a
noninteracting-system density by a configuration of one's choice.
Sahni {\em et al.} obtained the potentials using the differential
virial theorem \cite{hm}, whereas Harbola did so using the
constrained-search approach \cite{levy}. It is clear that because of the
multiplicity of potentials that could lead to a given density, one needs
an additional condition for mapping the density $\rho(\bf r)$ to a unique
potential $v(\bf r)$. For the ground-state density the Hohenberg-Kohn (HK)
theorem \cite{hk} fixes the Kohn-Sham (KS) system \cite{ks} uniquely - it
is that system where the lowest
energy orbitals are occupied.  For the excited-state density a
different criterion is needed. The issue of how to choose a unique
potential (KS system) for a given excited-state density has been
addressed earlier by Levy and Nagy (LN) \cite{ln}. They have
proposed a qualitative criterion, discussed below, for doing so. However, 
it has not been investigated quantitatively.

Furthering the work of \cite{smss} and \cite{harb}, Gaudoin and
Burke \cite{gb} have shown that even with a fixed configuration, one can
reproduce an excited-state density from more than one
potentials. They have worked with {\it non-interacting} fermions and  
have generated these potentials using the inverse linear
response of a system and have related the multiplicity of
these potentials to a property of the linear response kernel. Based on
this they have made two observations: (i) that one has to go
beyond the HK \cite{hk} theorem to understand the multiplicity of
potentials with the same density, and (ii) that the mapping
$\rho({\bf r})\rightarrow v({\bf r})$ is not unique. However, as
pointed out above, the uniqueness between a density and a potential exists
only for the ground-states. Thus the existence of more than one potential
for excited-states densities is not excluded by the HK theorem. And
precisely for this reason, an additional condition is needed to identify
one particular system as the KS system
representing an excited-state.

The interesting and thought provoking results of references
\cite{smss,harb,gb} have prompted us to pursue the matter of
density-to-potential mapping for excited-states further. Our
investigations in this direction form the contents of this paper. We show:
(i) that the constrained-search approach \cite{levy} itself is capable of
generating all potentials for noninteracting systems
giving the same density, thereby establishing once again the
importance of this approach \cite{ag1,ag2,ln,harb} in excited-state
density-functional formalism; (ii) in the examples taken, when the criterion 
proposed by LN \cite{ln} is applied to different noninteracting systems
corresponding to a given density, it correctly identifies the system that
should represent the excited-state density; and (iii) the
conclusion of Gaudoin and Burke \cite{gb} about the ``lack of HK
theorem for excited-states'' has been arrived at by taking into consideration
only the excited-state density $\rho(\bf r)$ and therefore do not apply to
the Levy-Nagy formalism. In the following, after describing
briefly the LN formulation of density-functional theory (DFT) for
excited states, we give examples of how its application gives a
unique $\rho({\bf r}) \rightarrow v({\bf r})$ map.

Like for the ground-state theory, the LN formulation \cite{ln}
provides a variational DFT approach for the $k$th excited state of an
$N$-electron interacting system by defining a unique universal
functional $F_k[\rho,\rho_0]$ such that the energy $E_k$ and the
density $\rho_k({\bf r})$ of this state are given by
\begin{eqnarray}
E_k &=& \min_{\rho \to N} \left\{ \int {\rm d}^3 r\,
v_{\text{ext}}({\bf r})\,\rho({\bf r}) + F_k[\rho,\rho_0]
\right\}\nonumber \\
 &=&  \int {\rm d}^3 r\ v_{\text{ext}}({\bf r})\,\rho_k({\bf r})
+ F_k[\rho_k,\rho_0]
 \label{1}
\end{eqnarray}
Here $v_{\text{ext}}({\bf r})$ is the external potential,
$\rho_0({\bf r})$ is the ground-state density of this system. Due to the
HK \cite{hk} theorem, $v_{\text{ext}}$ is a unique functional of $\rho_0$.
In the definition of the bi-density functional
\begin{equation}
F_k[\rho,\rho_0] = \min_{\Psi \to \rho,\;
\{\langle\Psi|\Psi_j\rangle=0,\;j<k\}} \langle \Psi | {\hat T} +
{\hat V}_{\text{ee}}| \Psi \rangle \;, \label{2}
\end{equation}
the $N$-electron trial wave function $\Psi$ belongs to the space
which is orthogonal to the space spanned by all lower $j$th-state
functions $\Psi_j$ of the system, $j < k$. ${\hat T}$ and ${\hat
V}_{\text{ee}}$ are the $N$-electron operators of the kinetic and
electron-electron interaction energies. In this formulation, the
mapping from the $k$th excited-state density $\rho_k({\bf r})$ to
the corresponding wave function $\Psi_k$ of the system follows from Eq.
(\ref{2}) after inserting there $\rho=\rho_k$ --- the minimizer in Eq.
(\ref{1}), because then $\Psi_k$ is the minimizer in Eq.
(\ref{2}). Since the lower states $\Psi_j$ ($j < k$) in Eq.
(\ref{2}) are determined by $v_{\text{ext}}[\rho_0]$, a functional of
$\rho_0$, it is clear that the ground-state density $\rho_0$
plays an important role in this DFT for excited states.

For each original, interacting system, one can introduce a
corresponding noninteracting system such that their densities are
the same. In the constrained-search approach, this is done by
minimizing the expectation value $\langle\Phi|\hat T|\Phi\rangle$, where
$\Phi$ is now a single Slater determinant of one-electron spin orbitals
that gives the density of interest. However, many different noninteracting
systems (potentials) can be related with the given
excited-state density $\rho_k({\bf r})$ of the interacting system. Of the
many $\Phi$'s (many systems) that may give the same density $\rho_k$, a
unique one, the KS system, is chosen by comparing the
ground-state densities of the corresponding noninteracting systems and the
true ground-state density $\rho_0$, and checking if the LN criterion for
identifying the Kohn-Sham system for an excited-state is satisfied. Thus,
let in a particular noninteracting system
[characterized by its potential $v({\bf r})$] the density of its
$m$th state, $\rho^v_m({\bf r})$, be the same as $\rho_k({\bf r})$. Its
ground-state density will be denoted accordingly as
$\rho^v_0({\bf r})$. Then the KS system connected with $\rho_k$ is
identified among the above noninteracting systems as the one whose
$\rho^v_0({\bf r})$ resembles $\rho_0({\bf r})$ most closely in a
least-squares sense. The LN \cite{ln} criterion intuitively defines the KS
system consistent with the adiabatic connection to the $k$th excited-state
of the interacting system (characterized in DFT by
$F_k[\rho,\rho_0]$). What it means is if the electron-electron
interaction in an interacting system is turned off slowly, keeping the
excited-state density unchanged, the corresponding ground-state density of
the resulting system will remain close to the true
ground-state density of the interacting system.  Thus of the many
noninteracting systems that give the same excited-state density, the one
whose ground-state density remains closest to the true
ground-state density of a given system is identified as the KS
system representing the excited-state of that system. The
noninteracting system so chosen should best resemble the true system
because, within the constraint of the equality of their
excited-state density, their ground-state densities match most
closely. This should also make their external potential resemble
each other by the HK theorem. We reiterate that in general a
particular density can be generated by a multitude of potentials;
and for each potential the associated energy functional is different
\cite{smss,harb} due to the difference in the noninteracting kinetic
energy of each system. However, to keep the structure of these
functionals and the corresponding potentials simple, it is important that
we have a criterion to choose one particular system. In this
connection we note that for the ground-state densities of
noninteracting electrons too, there exist \cite{smss,harb} more than one
noninteracting systems that give the same density $\rho_0$.
However, the one where the lowest energy orbitals are occupied, i.e.
$\rho^v_0({\bf r})=\rho_0({\bf r})$, is the chosen KS system, and it is
unique due to the HK theorem. For the excited-states the
uniqueness should be provided by the LN criterion. But, before applying it in
practice, this qualitative LN \cite{ln} criterion that
``$\rho^v_0({\bf r})$ resembles $\rho_0({\bf r})$ most closely in a
least-squares sense'' needs to be transformed into some quantitative form.

One of the ways that the difference between two densities $\rho_a$ and 
$\rho_b$ can be characterized quantitatively is by the squared distance 
in the functional space
\begin{equation}
\Delta[\rho_a,\rho_b] = \int_\infty {\rm d}^3 r\,\Bigl(\rho_a({\bf r})
-\rho_b({\bf r}) \Bigr)^2\,. \label{3}
\end{equation}
We propose to consider the value of $\Delta[\rho^v_0,\rho_0]$ as
representing the least-squares deviation of the density $\rho^v_0$ from
the density $\rho_0$. Then, applying the LN criterion, the
noninteracting system having the smallest $\Delta$ would be chosen as the
KS system. We are going to demonstrate on examples that the proposed
quantitative version of the LN criterion chooses the KS
system in agreement with intuitive expectations in the cases considered.
What it means is if the excitation corresponds (i) to a fixed external 
potential with several configurations(Sahni {\em et al.} \cite{smss} and Harbola
\cite{harb}) or (ii) to a particular configuration with different 
potentials (Gaudoin and Burke\cite{gb}), then the minimum deviation occurs 
only for the true configuration / for the exact potential.
It is also shown that if some particular, different quantitative measure 
of the distance between densities is used in the LN criterion, it may lead 
to erroneous choice of the KS system. Thus definition of an adequate 
quantitative form of the criterion is important and needs verification.

We now discuss the case of the excited-state density of a {\em noninteracting} 
Fermionic systems where the LN criterion is very transparent. This is
because in this case the difference between the ground-state density of
the KS system representing an excited-state and the true ground-state
density will be zero. Thus it is easily shown that the LN criterion fixes
the KS system to be the true system:  Suppose an excited-state density is
produced by two different potentials
\cite{gb}.  Each of these potentials (systems) has a unique
ground-state by the HK theorem.  Thus only that particular system which 
has the same ground-state density as the system under consideration can truly
represent the original system.  In the {\em interacting} system, the
condition of the ground-state densities being the same is replaced by the
condition of the proximity of the densities in the
least-square sense. We note that in the work of Gaudoin and Burke
\cite{gb} who have analyzed non-interacting systems only,
 this comparison with the ground-state density is not made. Hence
they raise a question which potential should be chosen from the
many available. It is clear from the discussions above that this
should be determined by the comparison of the ground-state densities of
the excited-states systems (which are connected with the given
$\rho_k$) with the true ground-state density of the system under
consideration. In the following we show through examples that this gives a
 KS noninteracting system which is consistent with the original
system not only in terms of their ground-states, but also in terms of the
configuration of the excited-states.

To construct, for an assumed electronic configuration, the
potentials leading to a given density through constrained-search
approach \cite{levy}, we have employed the procedure of Zhao and
Parr \cite{zp,zmp}. It produces the potential $v({\bf r})$ of a
noninteracting system by making the value of the system kinetic
energy $\sum_{i=1}^\infty n_i \langle \psi_i| -\frac{1}{2}\nabla^{2}
|\psi_i \rangle$ stationary with the constraint that the system
density $\sum_{i=1}^\infty n_i|\psi_i({\bf r})|^2$ equals the given
density $\rho_k({\bf r})$. Here, $\psi_i({\bf r})$ is the space
orbital, $n_i$ --- its occupation number, $\sum_{i=1}^\infty n_i = N$; any
configuration connected with a single Slater determinant can be
represented by a proper choice of $n_i$ from the numbers $0,1,2$.
Depending on the starting potential used to initialize this
procedure, various potentials can be generated from the input
$\rho_k$ for each assumed configuration. Thus this procedure is
capable of generating all possible systems (potentials) that
reproduce the density on hand. Evidently, the determined different
potentials give different ground-state densities (by the Hohenberg-Kohn
theorem \cite{hk}); the one which is the closest to the true ground-state
density singles out the KS potential of the excited-state DFT.

\begin{figure}[thb]
\includegraphics[width=3.0in,angle=0.0]{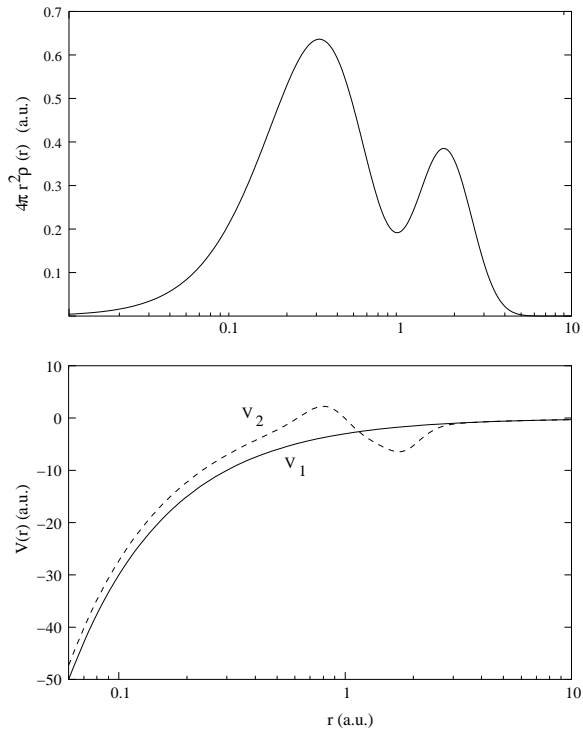}
\caption{Two potentials (lower panel) yielding the same excited state
density (upper panel) for $1s^1 2s^2$ state of a model Li atom.
Note that the x-axis scale in the upper and the lower panel is different.}
\label{niden-nipot}
\end{figure}

\begin{figure}[thb]
\includegraphics[width=3.0in,angle=0.0]{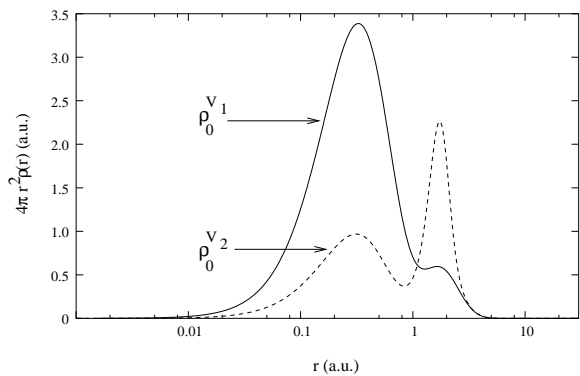}
\caption{Shown are the ground-state densities for the two potentials $v_1$
and $v_2$ of Fig.~\protect\ref{niden-nipot}.}
\label{nigden}
\end{figure}

As the first example we take a model Li atom: $N=3$ noninteracting
electrons moving in the potential $v(r)=-Z/r$, $Z=3$, resulting in the
hydrogen-like orbitals. Atomic units are used throughout.
The $k=1s^1 2s^2$ excited-state configuration of the model system
is considered. The potential $v_1(r)=v(r)$ and the alternative
potential $v_2(r)$ along with the excited-state density
$\rho(r)=\rho_k(r)$ are shown in Fig.~\ref {niden-nipot}.
Both potentials are generated using the Zhao-Parr \cite{zp,zmp} method
with the excited-state
density $\rho_k$ as the input. Since the
potentials $v_1$ and $v_2$ are different, their ground-state
densities can not be the same (by the HK theorem \cite{hk}).
Whereas the ground-state density corresponding to $v_1=-Z/r$ is the true
ground-state density of the system, that corresponding to $v_2$ should be
different --- they are shown in Fig.~\ref{nigden}. Indeed, the two
densities are dissimilar. If $v_2$ also were to represent
the KS system connected with the same excited-state density, the
difference in the ground-state densities for these two potentials
should be zero.

\begin{figure}[thb]
\includegraphics[width=3.0in,angle=0.0]{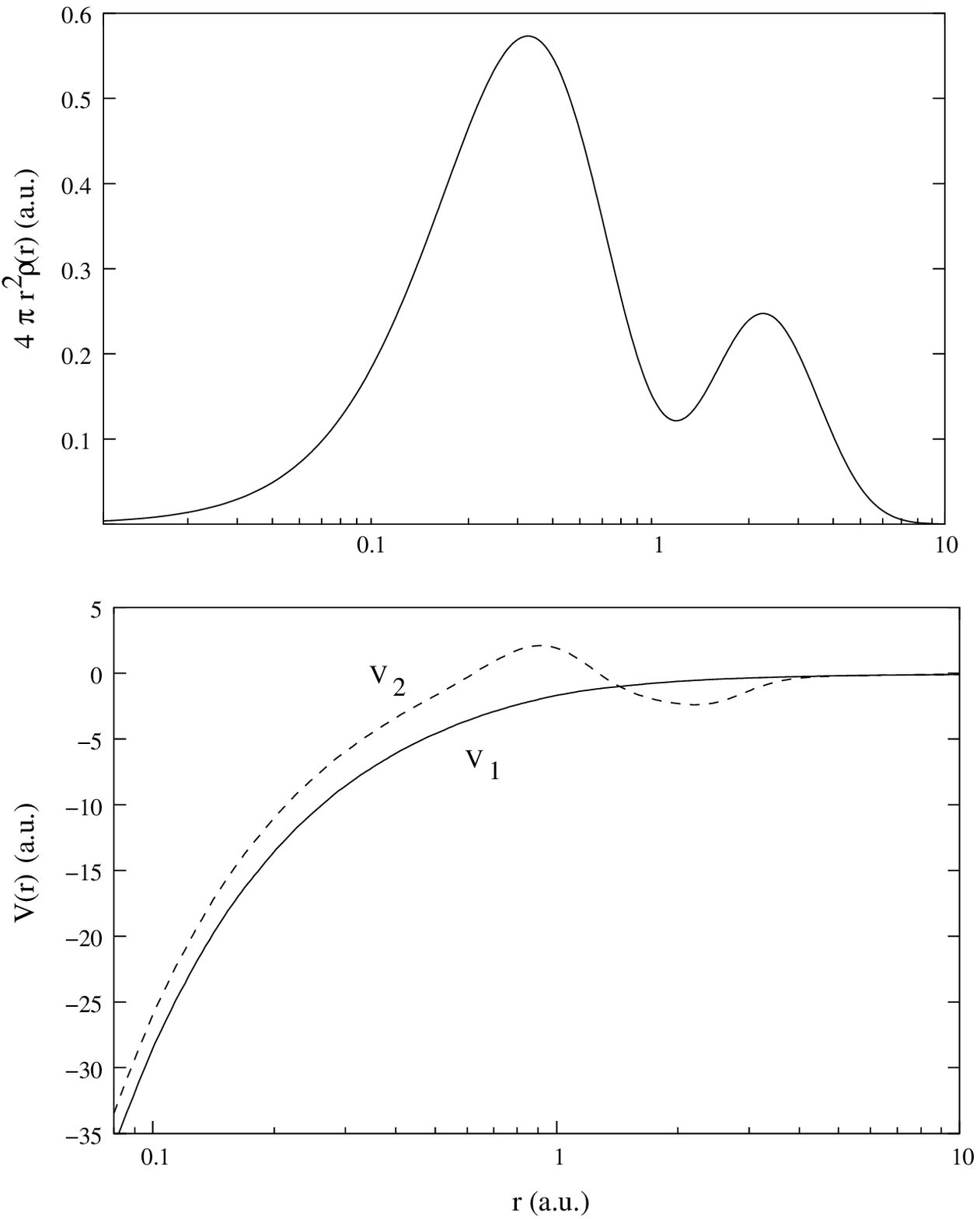}
\caption{Two potentials (lower panel) yielding the same excited state
density (upper panel) for the $1s^1 2s^2\,^2\!S$ state of the Li atom.
Note that the x-axis scale in the upper and the lower panel is different.}
\label{iden-ipot}
\end{figure}

\begin{figure}[thb]
\includegraphics[width=3.0in,angle=0.0]{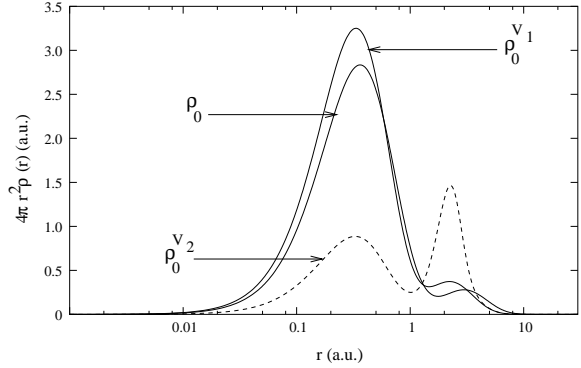}
\caption{Shown are the noninteracting ground-state densities for the two
potentials $v_1$ and $v_2$ of Fig.~\protect\ref{iden-ipot} along with the
interacting (exact) one.}
\label{g1-g2}
\end{figure}

As the second example we take the true Li atom: $N=3$ interacting
electrons moving in the external potential $v_{\text{ext}}(r)=-Z/r$,
$Z=3$, and consider its $k=1s^1 2s^2\,^2\!S$ excited-state
configuration. The density $\rho_k(r)$ of this state is represented by the
density calculated selfconsistently using the exchange-only
Harbola-Sahni(HS) \cite{hs} potential for this configuration. This density
is a good approximation to the exact solution and very close
\cite{sen,deb} to the Hartree-Fock density of this state. Shown in
Fig.~\ref{iden-ipot} are the two potentials $v_1(r)$ and
$v_2(r)$ reproducing the same density $\rho_k(r)$ as the $1s^1 2s^2$
excited-state densities of noninteracting systems; $v_1$ coincides with
the HS effective potential, $v_1(r)=-Z/r + v_{\text{H}}(r) +
v_{\text{x}}^{\text{HS}}(r)$. The constrained-search procedure,
mentioned earlier, was employed to generate both $v_1(r)$ and
$v_2(r)$. Although the excited-state densities of the two potentials are
the same, the ground-state densities of these potentials are
different --- that is what discriminates between the two potentials. In
Fig.~\ref{g1-g2} we plot the noninteracting ground-state
densities $\rho^v_0(r)$ for potentials $v=v_1$ and $v=v_2$, along
with  the interacting ground-state density $\rho_0(r)$. The latter is
obtained in the same approximation as applied for $\rho_k$,
namely with the HS \cite{hs} potential, now for the ground-state
$1s^2 2s$ configuration. As expected, the three densities are
different. However, the noninteracting ground-state density produced by
$v_1$ is quite similar to the ``exact'' HS ground-state density of Li. On
the other hand, that corresponding to $v_2$ is very
different from the ``exact'' one. We introduce also the potential
$v_3(r)$ (not shown), which reproduces $\rho_k(r)$ as the
ground-state density of a noninteracting system. This $v_3$ is
unique according to the HK theorem. Thus the density shown in the
upper panel of Fig.~\ref{iden-ipot} can also be labeled with
$\rho_{0}^{v_3}$ in analogy with Fig.~\ref{g1-g2}. To apply the LN
criterion, the squared distance, Eq.~(\ref{3}), between ground-state
densities is evaluated, giving $\Delta[\rho^v_0,\rho_0]=0.111, 1.467,
0.460$ for $v=v_1, v_2, v_3$, respectively. Thus, according to this
criterion, the KS system connected with the $k$th excited state of the Li
atom is given by the potential $v_1$. This result confirms
our intuitive expectation.

\begin{figure}[thb]
\includegraphics[width=3.0in,angle=0.0]{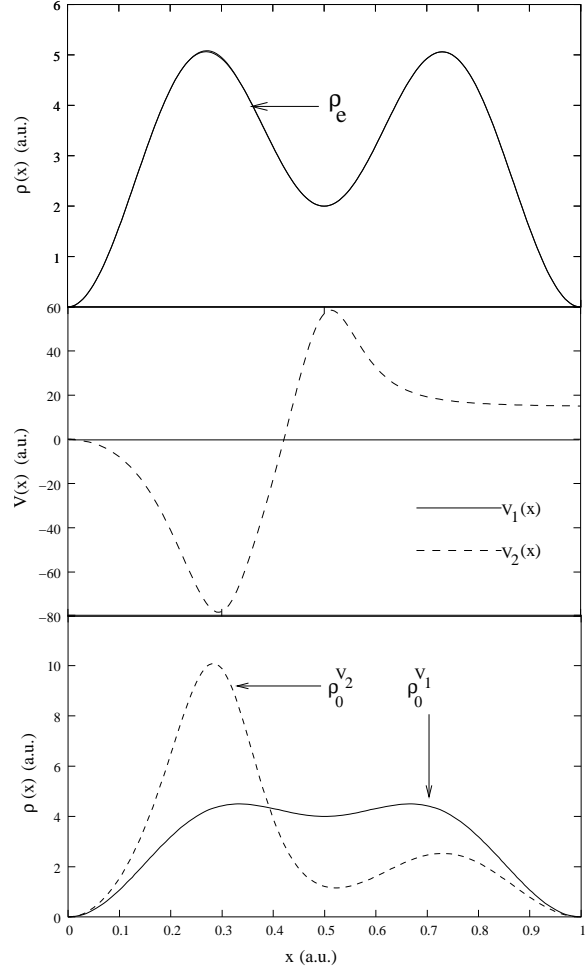}
\caption{Two potentials (middle panel) yielding the same excited-state
density (upper panel) along with their corresponding ground-state
densities (lower panel) for an excited state of the three-electron 1D
infinitely deep well model system.}
\label{den1-pot1}
\end{figure}

As the third example we take the He atom and consider its
$k=1s2s\,^1\!S$ excited-state, with the density $\rho_k$ taken from
\cite{cj}. This example was examined, in fact, in the previous work
\cite{harb} of one of us. Two potentials were obtained that
reproduce the density $\rho_k$: $v_1(r)$ as the density of the $1s 2s$
excited state of a noninteracting system, and $v_2(r)$ as the
density of the $1s^2$ ground state. To make use of the LN criterion, the
quantity $\bar\Delta[\rho^v_0,\rho_0] = \int_0^\infty {\rm d}r\, \{
\rho^v_0(r) - \rho_0(r) \}^2$ (for spherical densities) was
defined, and evaluated with $\rho_0(r)$, the ground-state density of the
true He atom, taken from \cite{kkt}. On the basis of the results
$\bar\Delta=0.273, 0.140$ for $v=v_1, v_2$, respectively, the author
concluded that the LN criterion \cite{ln} might not be proper for
finding the KS potential of the excited-state DFT. However, if the
distance between the ground-state densities is evaluated applying
the definition (\ref{3}) (proposed in the present paper), the result
$\Delta=0.086, 0.091$ for $v=v_1, v_2$, respectively, is obtained. Thus on
the basis of the LN criterion, we find that indeed the
$1s2s\,^1\!S$ state of He is properly represented by the KS system with
the potential $v_1$ that reproduces $\rho_k$ as the density of its $1s 2s$
configuration, in agreement with our intuition.

We note that although the considered examples seem to suggest that
comparing $\Delta$ as given by Eq.~3 gives KS system in accordance with 
the excited-states, Eq.~3 is not the only way of quantifying the LN
criterion. Better and more discriminating criteria may exist and should 
be looked for, particularly because of the limited number of examples 
considered in this paper and very small difference that is there for
the $He$ atom. But the emphasis in this paper is on showing how an
additional condition of comparing the ground-state densities may
lead to a proper choice of the KS system and that is shown amply by our 
work. Further work along these lines is in progress and will be reported 
in the future.

The above arguments apply equally well to the one-dimensional (1D) case
considered by Gaudoin and Burke \cite{gb}. Using another
conventional approach --- the van-Leeuwen and Baerends \cite{lb}
method --- for obtaining the noninteracting-system potentials, we
have reproduced not only the results of Gaudoin and Burke but have also
done many other calculations. The example we give in this paper is for the
following 1D model system: $N=3$ noninteracting electrons in an infinitely
deep box of unit length, in the excited state
obtained by putting one electron in the lowest-energy state and two
electrons in the second-lowest one. The excited-state density
($\rho_e$) and the corresponding two potentials that reproduce this
density in the same configuration are shown in Fig.~\ref{den1-pot1}. Also
shown are the ground state densities $\rho_0^{v_1}$ and
$\rho_0^{v_2}$ corresponding to these potentials. Again the
ground-state density given by the potential $v_2(x)$ is not the same as
that given by $v_1(x)$. Thus $v_2(x)$ can not represent the
excited-state density of the model.

To conclude, we have shown that if the ground-state density is
known, then the Levy-Nagy criterion may provide a proper map from an
excited-state density to the KS potential, $\rho({\bf r})
\rightarrow v({\bf r})$, provided the closeness of two densities is
adequately quantified, e.g., as in Eq.~(\ref{3}).  The criterion is 
exact for systems of {\em non-interacting} fermions. For {\em interacting}
electron systems, on the other hand, there could be other ways of defining 
this closeness quantitatively, but their effect should be verified 
(an example of failed definition was discussed). Our focus here is not 
on various definitions but rather how the proposed measure leads to a 
map from an excited-state density to the corresponding Kohn-Sham 
potential that is consistent with the configuration of excitation 
in the known cases.  Thus the excited-state energy can be expressed 
in terms of the density corresponding to this state and Kohn-Sham 
calculation can be done following the Levy-Nagy formulation.

{\bf Acknowledgment:} We thank Professor Viraht Sahni for his comments
on the manuscript.

\end{document}